# Abrupt Pulsation Resumptions in Deneb: An Update

*Joyce A. Guzik[1], Brian Kloppenborg[2], Noel Richardson[3], Jason Jackiewicz[4], Nancy Morrison[2,5], Tom Calderwood[2], and Andrzej Pigulski[6]*

[1]Los Alamos National Laboratory, Los Alamos, NM (joy@lanl.gov); [2]American Association of Variable Star Observers; [2]Embry Riddle Aeronautical University, Prescott, AZ; [4]New Mexico State University; Las Cruces, NM, [5]University of Toledo, Toledo, OH; [6]University of Wroclaw, Poland

**Subject Keywords**

Alpha Cygni Variables; stars: individual ($\alpha$ Cyg); AAVSO International Database; NASA TESS Spacecraft; BRITE Constellation Satellites

**Abstract**

Deneb, the prototype $\alpha$ Cygni variable, is a bright A2 Ia supergiant which shows irregular variability with a 12-day quasi-period, presumed to be caused by pulsations. At the 2023 AAVSO Annual Meeting we discussed radial velocity and photometry data from several sources showing that the 12-day variations begin abruptly at an arbitrary phase, damp out after several cycles, and resume at intervals of around 75 days. Additional data with more frequent time sampling and longer time series were needed to verify the existence and precision of the 75-day interval.

We have identified additional data sets and have intensified ground-based observing programs. Here we present analysis of 1) an 8.6-year photometric data set from the Solar Mass Ejection Imager; 2) BRITE Constellation light curves from six observing seasons of 60 to 180 days each, 2014-2021; 3) 4.6 years of radial velocity data from Morrison; 4) 1.4 years of radial velocity data from Eaton; and 5) additional V-band photometry from the AAVSO Photoelectric Photometry (PEP) section. Examining the SMEI data set, we find a most common 100 to 125 day interval between 'pulsation' resumptions. These resumptions sometimes skip intervals. We also find sudden large excursions in brightness and radial velocity which are distinct from the 'pulsation' resumptions and may or may not be data artifacts. We point out changes in the average level of Deneb's radial velocity which appear to be real given the accuracy of the measurements but are not explained.

**1. Introduction**

Despite being one of the brightest stars, Deneb (spectral type A2 Ia) holds many mysteries. Deneb is the prototype of the $\alpha$ Cyg variables, which are characterized by small-amplitude (0.1 mag) variations. Deneb varies in both brightness and radial velocity with a quasi-period around 12 days. These variations have been attributed to pulsations (see, e.g., Abt 1957; Lucy 1976), but, unlike typical pulsational variability, these variations do not have a regular period or amplitude[1]. Abt et al. (2023) found evidence in radial velocity data that these variations start abruptly at an arbitrary phase, damp out over a few cycles, and then resume at intervals of around 75 days. Sometimes there are also sudden excursions in brightness or radial velocity, which are distinct from resumption of the 12-day quasi-periodic variations.

---
[1] Further references to pulsation will be in quotation marks.

These analyses raise many questions, such as: How precise and regular is the time interval of around 75 days between resumption of the variations with quasi-period 12 days? Do these resumptions in fact occur abruptly and at an arbitrary phase? What is the cause of these variations? Why do they damp out and resume? Do other α Cygni variables show similar behavior? What is the evolutionary state of Deneb and the α Cygni variables?

At the 2023 AAVSO 112th Annual Meeting we discussed radial velocity data from Paddock (1935) and Abt (1957), photometry and radial velocity data from Richardson et al. (2011), photometry from the American Association of Variable Star Observers database (Kloppenborg 2023), and photometry by the NASA TESS spacecraft (Ricker et al. 2015) presented by Abt et al. (2023). (See also Guzik et al. 2023 AAVSO proceedings.). For the above-mentioned data sets, either the sampling was too sparse or the time series were too short to confirm the precision and repeatability of the suggested 75 day interval between 'pulsation' resumptions.

We have identified additional data and have begun new observing programs. Here we present analysis of an 8.6-year photometric data set from the Solar Mass Ejection Imager (Kloppenborg 2023), BRITE Constellation (Weiss et al. 2014) light curves, radial velocity measurements from Morrison extending the time series of Richardson et al. (2011), radial velocity measurements from Eaton (2020), and new photometry by the AAVSO Photoelectric Photometry (PEP) section.

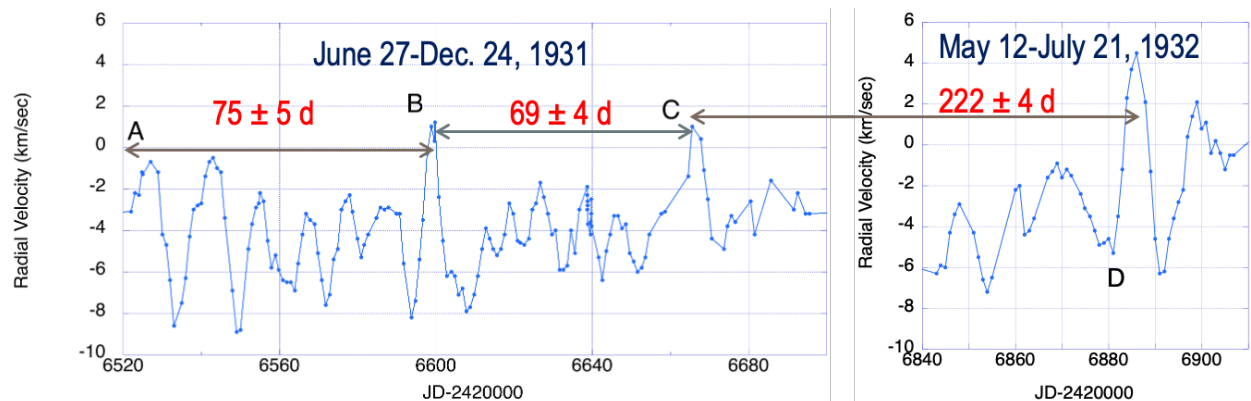

Figure 1. Radial velocity measurements from Paddock (1935) as presented by Abt et al. (2023). The interval between events C and D could correspond to 3 x 73-75 days.

## 2. Deneb Data Presented by Abt et al. (2023)

We first review observations from Abt et al. (2023) providing evidence for the 'pulsation' resumptions and large excursions in photometry and in radial velocity. Figure 1 shows radial velocity measurements by Paddock (1935). Intervals between points labeled A, B, and C are multiples of 69-75 days. The 'pulsation' resumption at A probably occurred a short time before the beginning of this time series. Also, there seem to be larger excursions without resumption of 'pulsations' at points B and C.

Figure 2 shows ground-based Deneb photometry and radial velocity measurements of Richardson et al. (2011) from April 1997 to December 2001. Several points are identified where large excursions are evident in one data set (E, F, H) and possibly in both data sets (G). There are three nearly equal intervals between events of 447-448 days, which could be a could be a multiple of a smaller interval, e.g., 6 x 74-75 days. Some events may have been missed because of data gaps; alternatively, perhaps the events are not regular and the star sometimes skips events.

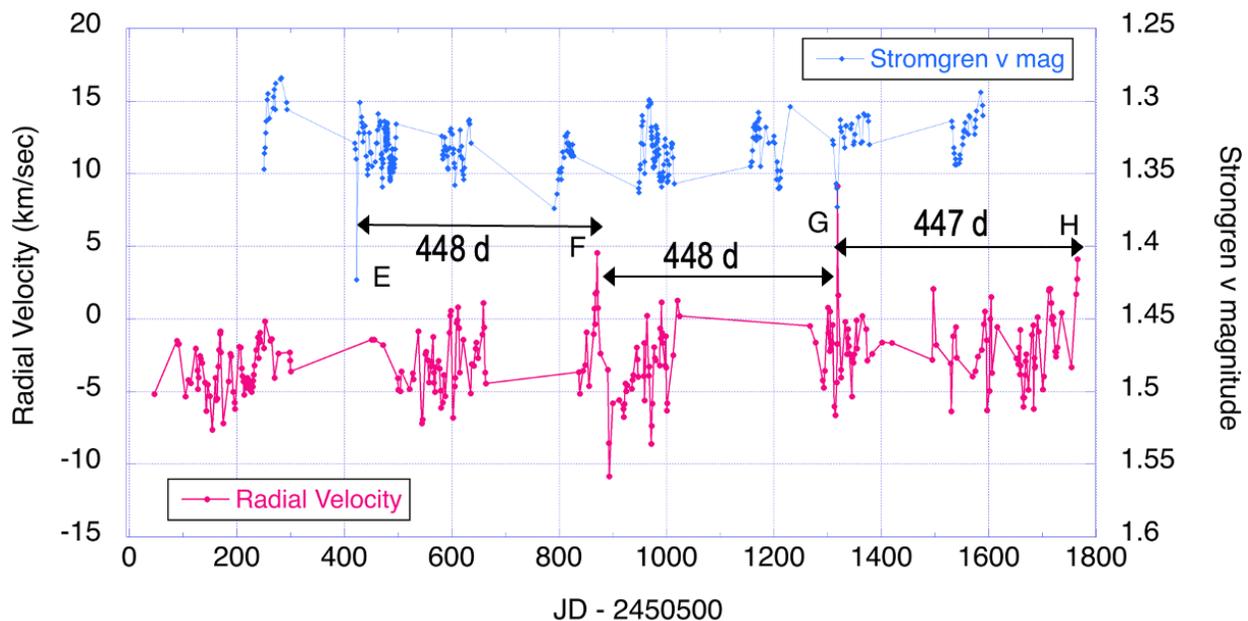

*Figure 2. Deneb photometry and radial velocity measurements from Richardson et al. (2011) as presented in Abt et al. (2023). Abrupt excursions in radial velocity and/or photometry occur at intervals of about 448 days.*

Deneb was in the field of view of the Transiting Exoplanet Survey Satellite (TESS, Ricker et al. 2015) during three 27-day observing 'Sectors' in 2021 and 2022. The 2-minute cadence high-level science product (HLSP) light curves are available at the Mikulski Archive for Space Telescopes (MAST, https://archive.stsci.edu). These light curves (Figure 3) show mostly irregular variations in Sector 41 and resumption of the quasi-periodic 12-day variations during Sector 55, continuing into Sector 56.

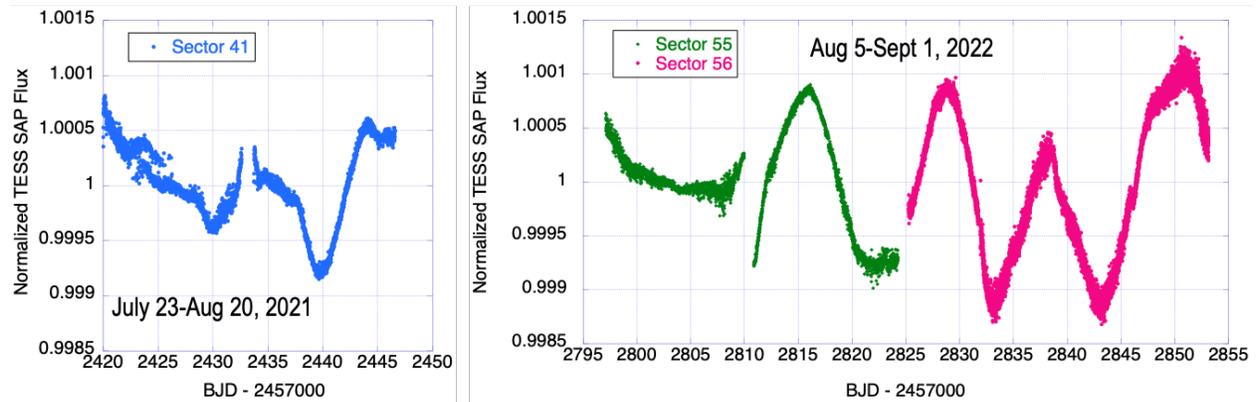

*Figure 3. Deneb photometry from TESS spacecraft as presented by Abt et al. (2023). Variations with quasi-period around 12 days resume around day 2810 in Sector 55 data.*

Abt et al. (2023) also examined Deneb V-magnitude photoelectric photometry (PEP) from the AAVSO International Database (AID) taken June 16, 2021–June 15, 2023 (128 data points). The PEP team has intensified its observing campaign for Deneb in 2024; the results are discussed in Section 3d.

**3. Additional Deneb Data**

After the AAVSO 112[th] Annual Meeting we identified and analyzed data from several sources: The Solar Mass Ejection Imager photometry, spanning 8.6 years; BRITE Constellation satellite photometry (2 to 6 month time series, 2014-2021); Morrison radial velocity data (4.3 years, 2002-2007); Eaton (2020) radial velocity data (1.4 years, 2008-2009); and AAVSO photoelectric photometry (PEP) data (118 new observations, June-October 2024).

*3.1 Solar Mass Ejection Imager Light Curves*

At the 112[th] AAVSO meeting, B. Kloppenborg informed us about Deneb photometric data taken by the Solar Mass Ejection Imager satellite (Jackson et al. 2004, Clover et al. 2011). This satellite was in a Sun-synchronous polar orbit from 2003-2011. Although the satellite was designed to study solar coronal mass ejections (CMEs), it also observed almost every bright (V < 6) star at a cadence of one data point per 102-minute (1.7-hour) Earth orbit. Kloppenborg processed these data to remove zero-point offsets and angle-dependent flux loss causing a 100-day curvature in the light curve.

We presented initial analysis of these data for the Society at Astronomical Sciences June 2024 meeting (see Guzik et al. 2024 SAS proceedings). Figure 4 shows the 8.6-year Deneb SMEI light curve. It is evident that additional processing is needed to remove remaining trends and offsets. Figure 5 zooms in on a few segments of the light curve. Sometimes the 12-day quasi-periodicity is evident, sometimes this variability almost disappears, and sometimes abrupt large discontinuities or excursions occur.

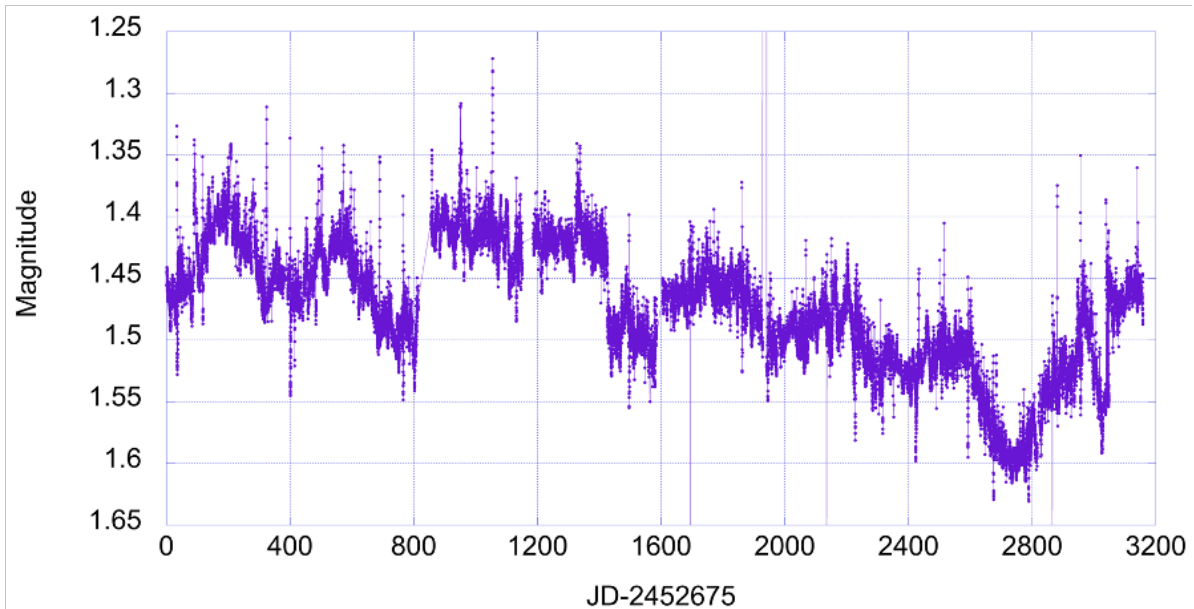

*Figure 4. Deneb light curve from Solar Mass Ejection Imager. These data require further processing to remove offsets and trends.*

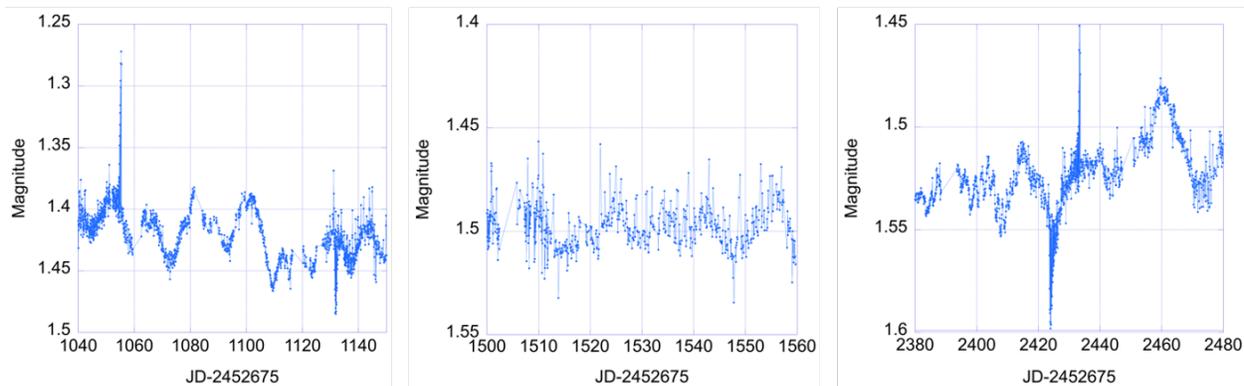

*Figure 5. Zoom-in on portions of Deneb SMEI light curve. Left: 12-day variations plus a couple of abrupt excursions. Center: Smaller amplitude variability, with 12-day variations less evident; Right: Larger amplitude variations and a large excursion.*

We reviewed the entire data set and measured the time intervals between resumptions of larger amplitude 12-day 'pulsation' variations, as well as the intervals between the abrupt excursions. Figure 6 (left) shows a histogram counting the number of events per time interval between resumptions of larger amplitude variations. The bin size is 50/4 = 12.5 days per bin. The resumptions occur most often at intervals of 100-125 days. These intervals are longer than the 75-day intervals hypothesized by Abt et al. (2023). The resumptions are not regular and appear to skip intervals. These resumptions also don't seem to occur abruptly at an arbitrary phase in the 'pulsation' cycle, as was suggested by Abt et al. (2023). Abrupt excursions/discontinuities in the Deneb SMEI data occur most often at intervals of 75-90 days (Figure 6, right). These events seem to be unrelated to the larger amplitude 'pulsation' resumptions. They may be caused by geomagnetic storm interference during the SMEI 102-minute polar orbit, and unrelated to Deneb. We discuss our initial attempts to confirm this hypothesis in Section 3c.

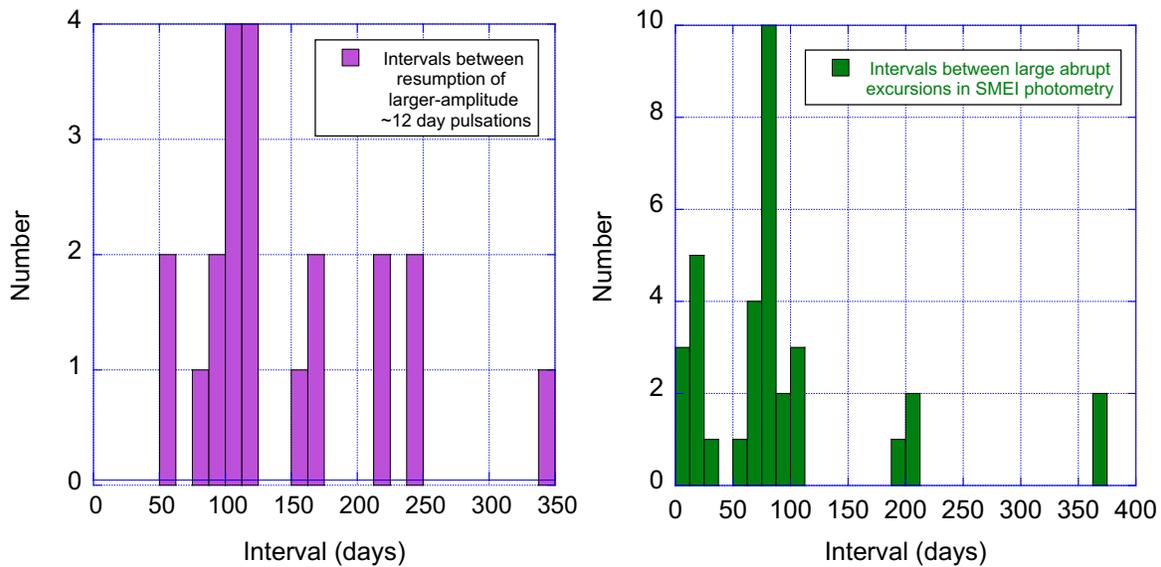

*Figure 6. Time intervals between resumption of larger-amplitude quasi-periodic 'pulsations' (left), and between abrupt excursions (right) in Deneb SMEI photometry.*

*3.2 BRITE Constellation Light Curves*

We also examined BRITE Constellation (Weiss et al. 2014) satellite data acquired during six observing seasons from 2014 through 2021. The BRITE Constellation consists of five nanosatellites in low Earth orbit, three observing in red and two in blue passbands. The satellites observe the target for 15-35 minutes per orbit, acquiring a photometric data point every 21 seconds, with 1 to 5-second exposures per observation. The BRITE data were decorrelated by A. Pigulski and some remaining trends/slopes were removed by J. Guzik.

The results of our analyses were first reported at the 2024 BRITE conference (Guzik et al. 2024; https://zenodo.org/records/14236753). Here we show a few BRITE light curves from 2015 and 2016 (Fig. 7). We found that 'pulsation' resumptions with amplitude around 20 millimag occur around days 1210, 1310 and 1600 after Barycentric Julian Date 2456000. The approximate 100-day interval between 'pulsation' resumptions at day 1210 and 1310 is consistent with the most common interval found in the SMEI photometric data. When Deneb was observed simultaneously by two BRITE satellites, one in red and the other in blue passband, the light curves show good correlation.

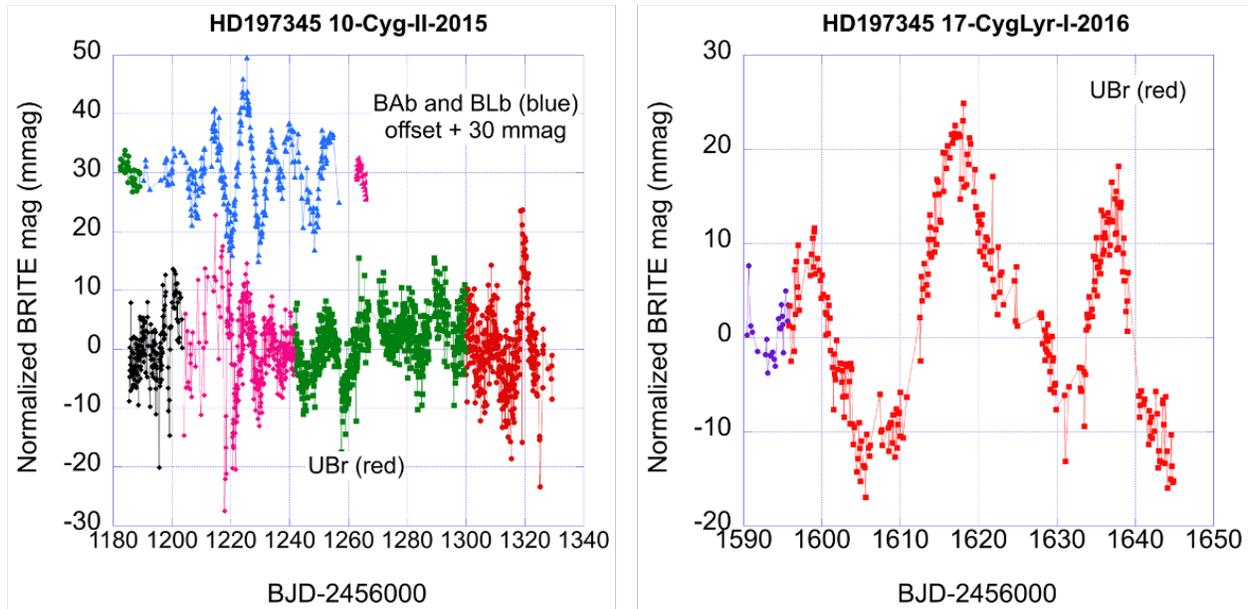

*Figure 7. Deneb BRITE Constellation photometry from 2015, showing 12-day variations in blue and red passbands (left), and large amplitude variations in 2016 in red passband (right)[2]. Larger amplitude (± 20 millimag) variations resume around days 1210, 1310, and 1600 after BJD 2456000.*

*3.3 Radial Velocity Measurements*

After the Richardson et al. (2011) project on Deneb photometry and radial velocity measurements concluded, high-resolution time-series spectra continued to be taken for several years at the Ritter Observatory 1-meter telescope of the University of Toledo. From these spectra, N. Morrison extracted 208 radial velocities covering the time period October 31, 2002 to January 16, 2007 (Fig. 8, left). A single spectral line, Si II λ6371, was used. The estimated uncertainty in a single measurement (mainly caused by the breadth of the line and by slight, time-variable profile asymmetries) is in the range 0.2 to 0.3 km/s.

While these data show variations consistent with the 12-day quasi-period, we also noticed average radial velocity level changes. For example, the average radial velocity is around -4 km/sec from day 600-800 after Heliocentric Julian Date 2452500, then increases to +1 km/sec around day 1000, and then decreases to -2 km/sec around days 1100-1200. We also notice smaller average radial velocity offsets in earlier radial velocity data sets; see, e.g., the data of Paddock (1935) and Richardson et al. (2011) of Figures 1 and 2. Morrison analyzed Ritter Observatory spectra for the radial velocity standard star β Oph and found it to be constant in radial velocity to ± 0.13 km/sec standard deviation during the same time frame. This result rules out an instrumental cause for the change in average velocity. There is no obvious periodicity to the level changes as would be expected if Deneb had a binary companion.

---

[2] The colors represent different segments or 'setups' of observations in a given field and are processed separately (Popowicz et al. 2017). The reason for splitting the data into setups is usually a change in observing conditions, and sometimes a change in exposure time.

Morrison also pointed out the availability of Deneb (HD 197345) radial velocity data from Eaton (2020), which we downloaded and plotted (Fig. 8, right). This data set included 134 observations from June 5, 2008 to November 3, 2009. The quoted uncertainty in radial velocity is 0.1 km/s. There is a possible large excursion event around day 1075 after HJD 2454000.

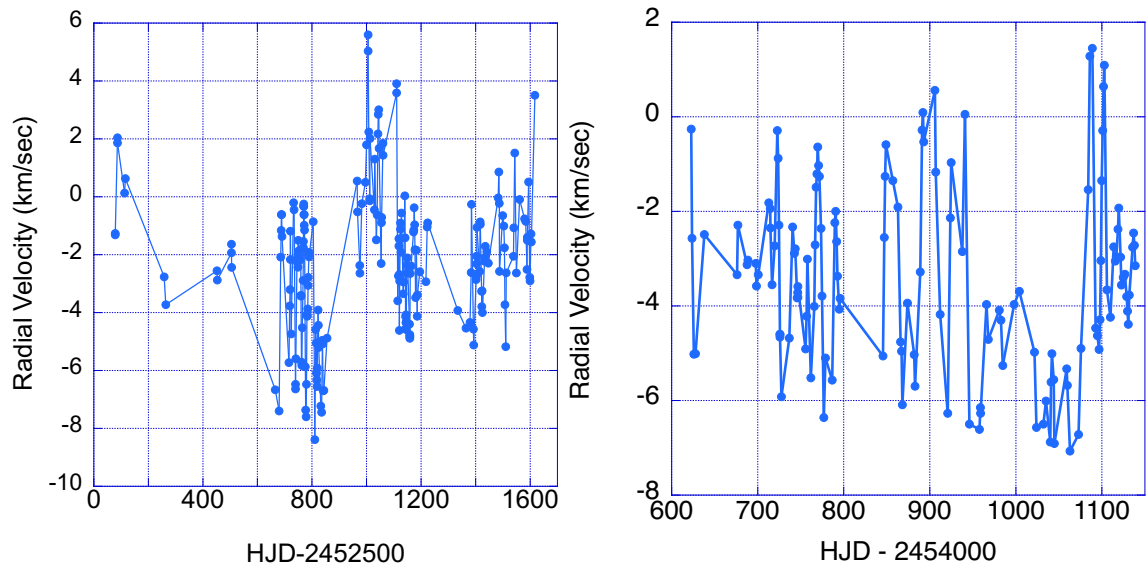

*Figure 8. Deneb radial velocity data from Morrison (left) and Eaton (2020, right). The average radial velocity appears to change in the Morrison data. A possible large excursion is seen in the Eaton data around HJD 2455075.*

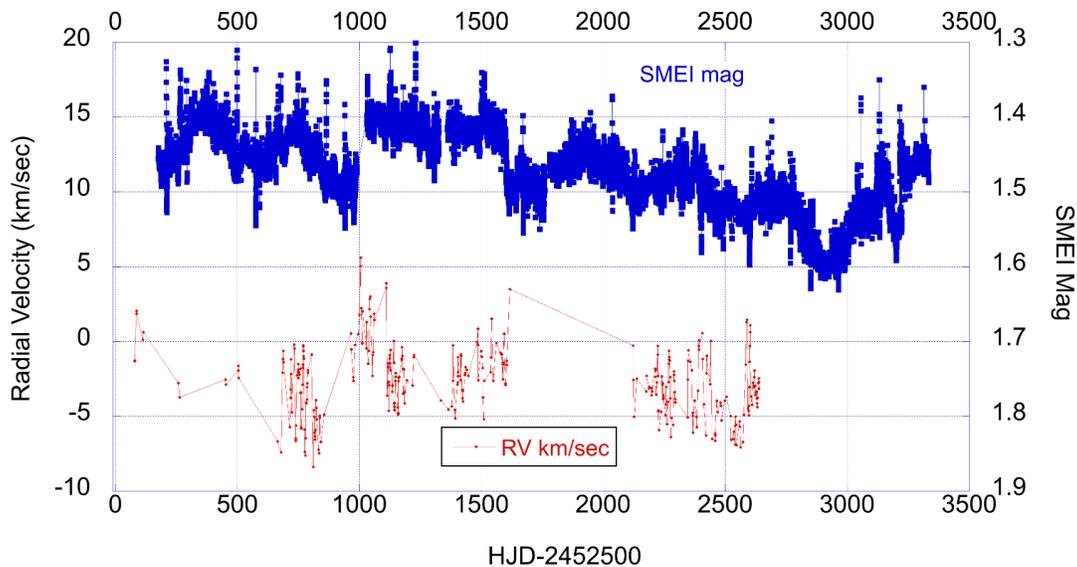

*Figure 9. Deneb Solar Mass Ejection Imager photometric data (blue) and Morrison/Eaton radial velocity data (red).*

The SMEI photometric data and Morrison/Eaton radial velocity data overlap in time. Figure 9 shows both data sets on the same plot. While more work will be required to identify a correlation between these data sets, we see that the large excursion in the Eaton radial velocity data is near in time to SMEI excursion of Figure 5 (right). Figure 10 zooms in on both data sets near this time.

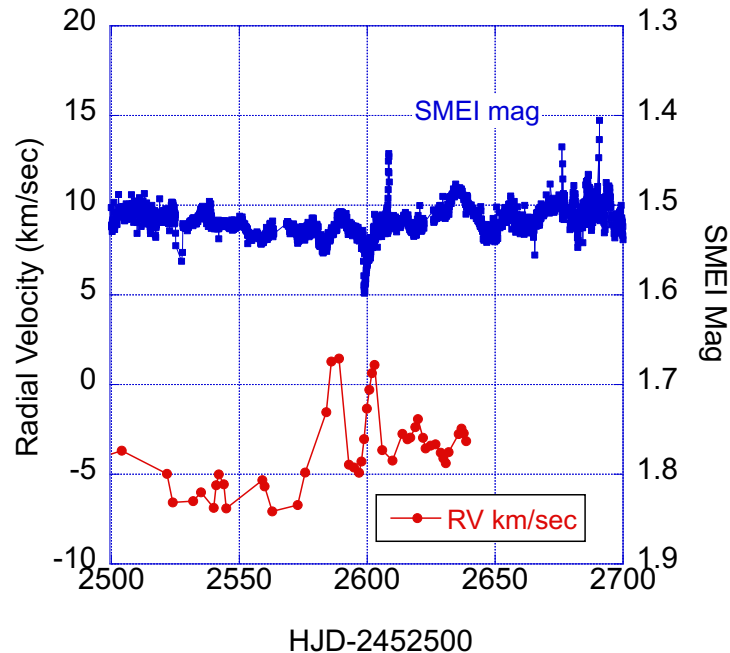

*Figure 10. Zoom-in on the large excursion event in SMEI data (Fig. 5, right) which is close in time to the large excursion in the Eaton (2020) radial velocity data (Fig. 8, right).*

We are working to determine the reality of the sudden excursions in the SMEI data, which were hypothesized to be caused by geomagnetic storms when the spacecraft passes through the Earth's magnetic field. We examined one variable and one 'constant' star very near to Deneb's coordinates: 55 Cyg = V1661 Cyg, with apparent magnitude 4.86, which is a blue supergiant also classified as an $\alpha$ Cyg variable; and 56 Cyg, a nearby main-sequence A-type star with apparent magnitude 5.06 believed to be a Hyades member. The light from these two stars would fall near to the same location on the SMEI cameras as Deneb's. If the excursions are caused by geomagnetic storms, the light curves of all three stars should show anomalies at the same time. Although we found that similar abrupt events are present in all three light curves, they do not correlate in time.

We also zoomed in further on a few of the abrupt excursions in the Deneb SMEI data and sometimes find oscillations every other data point (taken 1.7 hours apart) between values that are consistent with the light curve trends and points that deviate considerably. Such behavior, even if unexplained, seems even more likely to be artificial and should be cleaned from the light curves. However, we are reluctant to dismiss all of these events in the Deneb SMEI data as artifacts since large excursions are sometimes seen in other photometry and radial velocity data sets.

*3.4 AAVSO V-Band Photoelectric Photometry*

Since the 2023 AAVSO meeting, the AAVSO Photoelectric Photometry section has intensified its efforts to observe Deneb. At the 2024 AAVSO meeting, we presented PEP data from the AAVSO International Database (Kloppenborg 2023) from June 16, 2021 to October 16, 2024 (Fig. 11, 246 data points). Eleven observers with observing codes CTOA, FBA, FXJ, MPFA, BVE, DFR, BWU, GMV, GTIA, WIG, and DJAD contributed V magnitude observations. The median uncertainty of the data points is 0.003 mag. At the 2023 meeting, we noted that a large excursion may have been caught around day 200 after JD 2459300 (Fig. 11, left). There may be another event just before day 1200, but the coverage is a little too sparse to confirm. After day 1200 there are irregular variations with periods mostly longer than 12 days (Fig. 11, right).

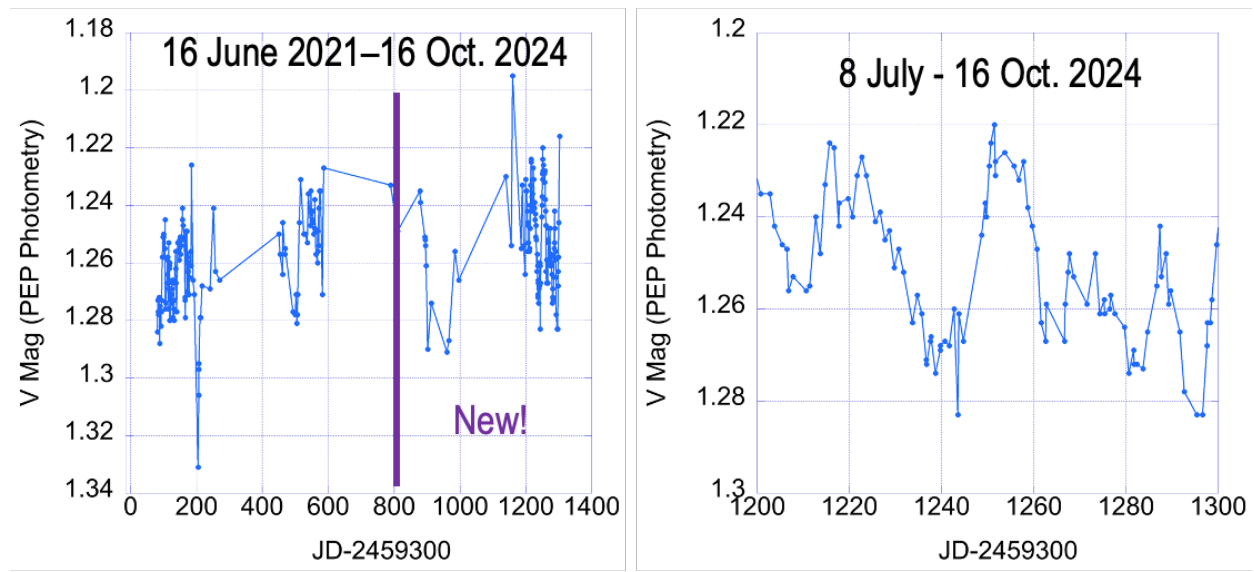

*Figure 11. Left: Deneb V magnitude photometric data from AAVSO International database (AID, Kloppenborg 2023). Data examined after the 2023 AAVSO meeting begin at day 800 after JD 2459300. Right: Zoom-in on new data, where at least one data point was acquired nearly every night. The median uncertainty in the data points over the time span of each figure is 0.003 magnitudes.*

**4. Summary and Future Work**

To summarize highlights from new data examined:

*SMEI Photometry:* Large-amplitude 'pulsation' resumptions occur most commonly every 100-125 days, longer than the 75-day intervals hypothesized by Abt et al. (2023). Sometimes intervals are shorter or are skipped entirely. Abrupt excursion events occur most commonly every 75 to 90 days. Most of these are likely to be data artifacts, but we haven't been able to prove that geomagnetic storms are responsible for these events, as hypothesized.

*BRITE Photometry*: We find one 100-day interval between large-amplitude 'pulsation' resumption, consistent with the most common SMEI interval.

*Radial Velocity:* Average radial velocity changes were noticed in recently reduced data from Morrison. One possible large excursion is found in data from Eaton (2020). These data sets overlap in time with SMEI photometry, so correlations can be studied. One of the SMEI large excursion events seems to correlate with the Eaton radial velocity event.

*AAVSO Photometry*: High-quality, high-cadence data are being acquired by the AAVSO Photoelectric Photometry section.

For the future, we will clean up the data for Deneb and other bright stars observed by SMEI, attempting to understanding and remove artifacts. Cotton et al. (2024) find that Deneb is a polarimetric variable, with an event occurring just after 'pulsations' resumed in TESS data. We are looking into whether amateurs can carry out polarimetry observations of Deneb and other $\alpha$ Cyg variables. We will look for additional TESS data for Deneb. TESS observed Deneb during Sector 76 in 2024, but a HLSP light curve was not yet available at MAST. In addition to continuing to observe Deneb, the AAVSO PEP section is also intensifying its observations for other $\alpha$ Cyg variables, such as Rigel.

In answer to a question at the 113th AAVSO meeting, Morrison noted that slight line-profile variations are evident in the Deneb spectra. These may provide important clues to the nature of Deneb's variability.

We are also beginning to use stellar modeling to investigate the evolution and pulsational stability of massive stars near the location of Deneb in the Hertzsprung-Russell diagram. A poster on this work was presented by A. Moore at the 2024 AAVSO meeting (see proceedings paper).

We hope that these studies will lead to a better understanding of the origin of the variability and the evolutionary state of Deneb and the $\alpha$ Cygni variables.

**Acknowledgements**

We acknowledge with thanks the variable star observations from the AAVSO International Database contributed by observers worldwide and used in this research. We also acknowledge the students and staff of Ritter Observatory (The University of Toledo), who acquired the spectra that we measured for radial velocity but who are too numerous to list here. This collaboration was facilitated by a Los Alamos National Laboratory Center for Space and Earth Sciences grant XX8P SF2. J.G. acknowledges support from Los Alamos National Laboratory, managed by Triad National Security, LLC for the U.S. DOE's NNSA, Contract #89233218CNA000001. Based on data collected by the BRITE Constellation satellite mission, designed, built, launched, operated and supported by the Austrian Research Promotion Agency (FFG), the University of Vienna, the Technical University of Graz, the Canadian Space Agency (CSA), the University of Toronto Institute for Aerospace Studies (UTIAS), the Foundation for Polish Science & Technology (FNiTP MNiSW),

## References

Abt, H.A. 1957, The Variability of Supergiants, ApJ 126, 138

Abt, H.A., Guzik, J.A., and Jackiewicz, J. 2023, The Abrupt Resumptions of Pulsations in α Cygni (Deneb), PASP 135, 1054, 124201

Clover, J.M., et al. 2011, Epsilon Aurigae light curve from the Solar Mass Ejection Imager, AAS Meeting Abstracts 217, 257.02

Cotton, D. V., et al., 2024, Deneb is a large-amplitude polarimetric variable, ApJ Letters 967, L43

Eaton, J.A., 2020, 35,000 Radial Velocities for 348 Stars from the Tennessee State University Automatic Spectroscopic Telescope, JAAVSO 48, 91

Guzik, J.A., Abt, H., Jackiewicz, J., and Kloppenborg, B.K. 2023, Abrupt Periodic Pulsation Resumptions in Deneb, Proceedings of the 112th Annual Meeting of the AAVSO, https://www.aavso.org/112, doi: 10.48550/arXiv.2410.23936

Guzik, J.A., Kloppenborg, B., and Jackiewicz, J. 2024, Deneb and the alpha Cygni Variables, Society for Astronomical Sciences Symposium on Telescope Sciences 2024 proceedings, eds. J.C. Martin, R.K. Buchheim, R.M. Gill, W. Green, and J. Menke, https://socastrosci.org/wp-content/uploads/2024/08/2024-Proceedings_Ver1.3d.pdf, doi: 10.48550/arXiv.2410.23985

Guzik, J.A., Richardson, N.D., Kloppenborg, B., Jackiewicz, J., and Pigulski, A. 2024, Deneb's Variability as Viewed by BRITE Constellation and the Solar Mass Ejection Imager, The BRITE Side of Stars Electronic Proceedings, University of Vienna, August 20-23, 2024, Online at https://britestars.univie.ac.at/home/, id.68, doi: 10.5281/zenodo.13858241

Jackson, B.V., et al. 2004, The Solar Mass Ejection Imager (SMEI) Mission, Sol. Phys. 225, 177

Kloppenborg, B.K. 2023, Observations from the AAVSO International Database, https://www.aavso.org

Lucy, L.B. 1976, An analysis of the variable radial velocity of alpha Cygni, ApJ 206, 499

Paddock, G.F. 1935, Spectrographic Observations of Alpha Cygni, Lick Obs. Bull. 17, No. 472, 99

Popowicz, A., et al. 2017, BRITE Constellation: data processing and photometry, A&A 605, A26

Richardson, N.D., Morrison, N.D., Kryukova, E.E., and Adelman, S.J. 2011, A five-year spectroscopic and photometric campaign on the prototypical alpha Cygni variable and A-type supergiant star Deneb, AJ 141, 17

Ricker, G.R., et al. 2015, Transiting Exoplanet Survey Satellite, Journal of Astronomical Telescopes, Instruments, and Systems, Volume 1, id. 014003

Weiss, W., et al. 2014, BRITE-Constellation: Nanosatellites for Precision Photometry of Bright Stars, PASP 126, 573